\documentclass[aps,prd,preprint,superscriptaddress,tightenlines,nofootinbib]{revtex4}
\usepackage{graphicx}% Include figure files
\usepackage{dcolumn}% Align table columns on decimal point
\usepackage{bm}% bold math
\begin{document}
\newcommand{\Ufours}{ \Upsilon$(4S)$ }
\newcommand{\Uths}{ \Upsilon{\rm (3S)} }
\newcommand{\Utws}{ \Upsilon$(2S)$ }
\newcommand{\Uons}{ \Upsilon$(1S)$ }
\newcommand{\Ums}{ \Upsilon$(mS)$ }
\newcommand{\Uns}{ \Upsilon$(nS)$ }
\newcommand{\chibpr}{\chi_{b}^{\prime}}
\newcommand{\piz}{\pi^{0}}
\newcommand{\gev}{\ \rm GeV}
\newcommand{\gevoc}{\ \rm GeV/c}
\newcommand{\mev}{\ \rm MeV}
\newcommand{\mevoc}{\ \rm MeV/c}
\newcommand{\thtwdec}{ \Upsilon$(3S) $\to \pi^+ \pi^- \Upsilon {\rm (2S)} }
\newcommand{\signal}{ \chi_{b}$(2P)$\to\pi^{+}\pi^{-}\chi_{b} {\rm (1P)} }
\newcommand{\altsignal}{ \chi_{b}^{\prime}\to\pi^{+}\pi^{-}\chi_{b} }
\newcommand{\signalneut}{ \chi_{b}$(2P)$\to\pi^0 \pi^0\chi_{b} {\rm (1P)} }
\newcommand{\neutsignal}{ \chi_{b}^{\prime}\to\pi^0 \pi^0\chi_{b} }
\newcommand{\signalboth}{ \chi_{b}$(2P)$\to\pi\pi\chi_{b} {\rm (1P)}  }
\newcommand{\bothsignal}{ \chi_{b}^{\prime}\to\pi\pi\chi_{b} }
\newcommand{\lowerdecay}{\chi_{b}\to\gamma\Upsilon {\rm (1S)} }
\newcommand{\upperdecay}{\Upsilon$(3S)$\to\gamma_{1}\chi_{b}{\rm (2P)} }
\newcommand{\onellpair}{\Upsilon$(1S)$\to\ell^{+}\ell^{-}}
\newcommand{\chigg}{\chi^{2}_{\gamma\gamma}}
\newcommand{\chiups}{\chi^{2}_{\Upsilon}}

\preprint{CLNS 05/1937}       % for CLNS notes
\preprint{CLEO 05-25}         % for CLNS notes
\title{Experimental Study of 
$\chi_{b}$(2P)$\to\pi\pi\chi_{b}$(1P)}
%-------- INSERT HERE ------------
\author{C.~Cawlfield}
\author{B.~I.~Eisenstein}
\author{I.~Karliner}
\author{D.~Kim}
\author{N.~Lowrey}
\author{P.~Naik}
\author{C.~Sedlack}
\author{M.~Selen}
\author{E.~J.~White}
\author{J.~Williams}
\author{J.~Wiss}
\affiliation{University of Illinois, Urbana-Champaign, Illinois 61801}
% ...
%\author{(CLEO Collaboration)} %FOR PRD_SPECIAL_CHANGEME
\collaboration{CLEO Collaboration} %FOR PRL,CLNS
\noaffiliation

%-------- END INSERT ------------

%please hard code the date when you have a final draft and submit to CLEOAC
%\date{\today}
\date{November 7 2005}

\begin{abstract} 
We have searched for the di-pion transition 
$\chi_{b}$(2P)$\to\pi\pi\chi_{b}$(1P)
%$\signalboth$ 
in the CLEO III sample of $\Uths$ decays
in the exclusive decay chain:
$\Uths\to\gamma\chi_{b}$(2P), 
$\chi_{b}$(2P)$\to\pi\pi\chi_{b}$(1P),
%$\signalboth$,
$\chi_{b}$(1P)$\to\gamma\Upsilon$(1S), 
%and, 
%finally, 
$\onellpair$.
Our studies include both $\pi^{+}\pi^{-}$ and
$\piz\piz$, each analyzed both in fully reconstructed
events and in events with one pion undetected.
We show that the null 
hypothesis
is not substantiated. 
%The di-pion invariant mass distribution is shown for the charged decay $\signal$. 
Under reasonable 
assumptions, we find the
partial decay width to be 
$\Gamma(\chi_{b}$(2P)$\to\pi\pi\chi_{b}$(1P)$) =   
%= 3\cdot\Gamma_{\piz\piz} 
%= \frac{3}{2} \Gamma_{\pi^{+}\pi^{-}}=
(0.83 \pm 0.22 \pm 0.08 \pm 0.19)~{\rm keV},$
with the uncertainties being statistical, internal 
CLEO systematics, and common systematics from outside
sources.
\end{abstract}

\pacs{13.25.Gv} %hadronic decays of onia
\maketitle

% Insert body of the text here.
\section{Introductory Material}
\label{sec:intro}
Heavy quarkonia, either $c\overline{c}$ or $b\overline{b}$,
have provided good laboratories for the study of the strong
interaction.  New, large data samples at CLEO/CESR and 
BES/BEPC have renewed the interest in heavy quarkonia\cite{QWG}.

Although copiously produced in
electric dipole (E1) transitions from the $\Uths$ and $\Utws$, the 
$\chibpr$ ($2^{3}P_{J^{\prime}}$) 
and $\chi_{b}$ ($1^{3}P_{J}$) 
are largely unexplored.
%have very little known about them.  
The dominant hadronic transitions among the heavy quarkonia
%($Q\bar{Q} = c\bar{c}, b\bar{b}$) 
involve di-pion emission,
characterized by Yan\cite{Yan}
%as shown in Fig.~\ref{fig:soft}, 
as the emission
of two soft gluons which then hadronize as a di-pion system.
These have been studied for transitions among the quarkonia 
$^{3}S_{1}$
states, but have not been observed 
%(yet) 
in other quarkonia
transitions
such as $\eta_{c}^{\prime}\to\pi\pi\eta_{c}$ or the $\chibpr$
decays, which are the subject of this work.  
% RSG 16Sep05
% This section removed in that we no longer evaluate mpipi.
%Of particular interest in such
%transitions is the distribution of di-pion invariant mass ($m_{\pi\pi}$).
%%distribution.  
%The $m_{\pi\pi}$ spectra for the
%decays $\psi^{\prime}\to\pi\pi J/\psi$~\cite{BES},
%$\Uths\to\pi\pi\Utws$, and $\Utws\to\pi\pi\Uons$~\cite{CLEO2Sto1S} 
%%all seem to 
%agree with the Yan model\cite{Yan}.
%The situation is quite different
%for the decay $\Uths\to\pi\pi\Uons$\cite{CLEO3Sto1S87,CLEO3Sto1S94},
%%which has the largest available
%%energy ($Q$) of any of these transitions.  
%for which the 
%spectrum of $m_{\pi\pi}$ has two peaks, one at low invariant mass
%and one at high invariant mass.  
% RSG 16Sep05

New interest in $\chibpr$ decays has also been generated by the 
CLEO observation\cite{omegaPRL} of a large branching fraction
for the decay
%New interest given large decay rate for 
$\chibpr\to\omega\Uons$.  This is the only presently known
hadronic decay of 
the $P$-wave $b\bar{b}$ states and the only hadronic bottomonium
transition that is not through $\pi\pi$.    

\begin{figure}[thb]
\centerline{\includegraphics[width=8.5cm]{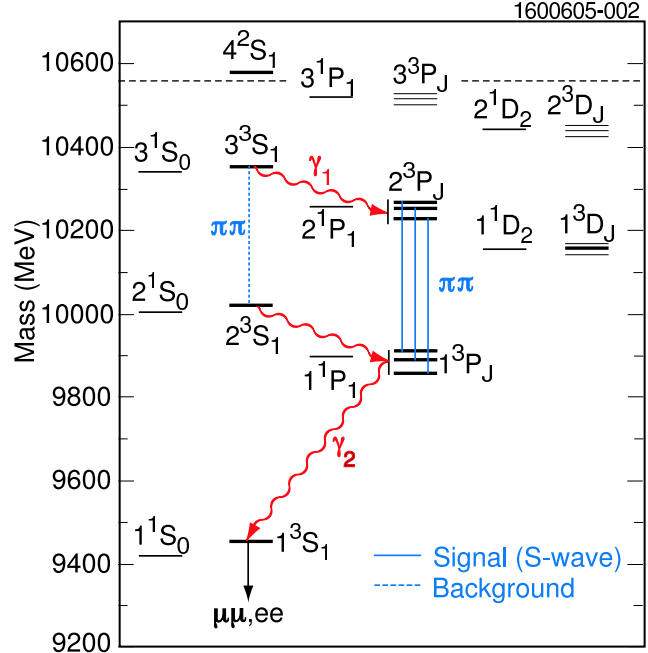}}
\caption{The decay process under study and the
main background process, denoted 
in the text
as
``$\pi\pi~\gamma~\gamma$''.  Note that these have the same
$\gamma_{2}$, so that the energy of this photon is not a 
distinguishing observable.}
%and the same number of photons, pions and leptons. }
\label{fig:signal_and_TS} 
\end{figure}

We have investigated another hadronic transition
of the 
$\chibpr$, namely 
%$\signalboth$.  
$\bothsignal$.  
As shown in 
Fig.~\ref{fig:signal_and_TS},
this search starts with the E1 transition
$\Uths\to\gamma_{1}\chibpr$, 
%then has 
followed by 
the signal process
%$\signalboth$, 
$\bothsignal$, 
and the 
%daughter 
resulting
$\chi_{b}$ decay (again via
an E1 transition) 
as $\chi_{b}\to\gamma_{2}\Uons$ with $\onellpair$.
Thus the final state has two photons, two low-momentum (``soft'')
%charged 
pions and two high-momentum leptons.  
In this 
%Letter/Article 
Article
we
(i) establish 
%the observation of 
this $\chibpr$ decay and, 
%(ii) present the observed $m_{\pi^{+}\pi^{-}}$ spectrum, and,
with reasonable assumptions, 
%(iii) determine the partial width $\Gamma_{\pi\pi}$.
(ii) estimate the partial width 
$\Gamma_{\pi\pi} \equiv \Gamma(\chi_{b}$(2P)$\to\pi\pi\chi_{b}$(1P)$)$.

The main background to our signal, also shown in
Fig.~\ref{fig:signal_and_TS},
has $\Uths\to\pi\pi\Utws$, followed by an E1 cascade 
through the $\chi_{b}$ states 
to the
$\Uons$.  This background process, which we will denote as
``$\pi\pi~\gamma~\gamma$'', has the same number of pions, leptons
and photons, with similar kinematics.  While this 
means we need
stringent selection criteria to define the signal, it also 
provides
%affords
a known process 
with a nearly identical final state 
against which to test our analysis procedures.

The data were collected at the Cornell Electron Storage Ring
using the CLEO III\cite{Viehhauser} detector configuration.  The components
most critical for this analysis were the Cs-I electromagnetic calorimeter and 
the charged particle tracking system, each covering $\sim$93\% of 
the $4\pi$ solid angle.  
Consisting of 7800
crystals, the calorimeter was originally installed in the 
CLEO~II configuration\cite{Kubota},
with some reshaping and restacking for CLEO~III
to allow more complete solid angle coverage.
%over the entire solid angle. 
The shower energy resolution,
$\sigma_{E}/E$,
is 4\% at 100 MeV and 2\% at 1 GeV in the barrel
region, defined as $|(\cos \theta)| < 0.80$, with
$\theta$ the dip angle with respect to the beam axis.
Complemented at small radius by a 4-layer double-sided
silicon vertex detector,
a new drift chamber\cite{Peterson}
was installed for CLEO III; 
its endplate design minimizes material,
enhancing the resolution of the endcap
electromagnetic calorimeter, extending the
solid angle coverage to $|(\cos \theta)| \sim 0.93$.

The signal was searched for in 1.39 fb$^{-1}$ of data accumulated at
the 
center of mass energy corresponding to the 
$\Uths$ resonance, consisting of $(5.81 \pm 0.12)\cdot 10^{6}$
resonance decays\cite{1D}.  We also used 8.6 fb$^{-1}$ of data taken
at $\sqrt{s}\approx 10.56$ GeV (``high-energy continuum'') and 0.78
fb$^{-1}$ of data taken at the $\Utws$, or roughly
$5.5\cdot 10^{6}$ decays\cite{Artusoetal}, to study and evaluate
backgrounds.

We used large Monte Carlo simulations based on 
{\tt GEANT3.211/11}\cite{GEANT}
to estimate
our efficiencies and tune our selection criteria.  In addition
to the signal process, we simulated: 
(i) the main background
process, ``$\pi\pi~\gamma~\gamma$'', as described above and
in Fig.~\ref{fig:signal_and_TS};
(ii) $\Uths\to\pi\pi\Utws$ with $\Utws\to\ell^{+}\ell^{-}$, a
process with higher statistics and similar pion kinematics, to
help confirm our efficiency determinations;
(iii) $\Uths\to\pi\pi\Utws$ with $\Utws\to\pi^{0}\pi^{0}\Uons$,
which could mimic our signal multiplicity if two photons were missed;
(iv) ``generic'' $\Uths$
Monte Carlo, which uses all known properties and modes of $\Uths$
decay, but for which the backgrounds (i) through (iii) are tagged
and not analyzed;
(v) $q\overline{q} (q = u,d,s,c)$ 
% corrected typo in list of quarks - RSGalik - 04Jan06
continuum processes at the $\Uths$ center of mass energy; 
(vii) for the charged pion decay channel,
$\Uths\to\gamma\chibpr$ with $\chibpr\to\omega\Uons$ and
$\Uons\to\ell^{+}\ell^{-}$, which has the same initial photon transition
as our signal but would have an additional photon in the decay
of the $\pi^0$ resulting from the $\omega$ decay to
$\pi^{+}\pi^{-}\pi^{0}$; and
(viii) for the neutral pion decay channel,
$\Uths\to\eta\Uons$ with $\eta\to\piz\piz\piz$;
%which has two more
%pions than the signal process (which could have been missed by the
%detector); 
we used ${\cal B}(\Uths\to\eta\Uons) = 2.2\cdot 10^{-3}$,
which is the present 90\% CL upper limit for this decay.

In our studies we assumed that there were no $D$-wave contributions to the
decays, only $S$-wave, so that $J^{\prime} = J$.  This assumption is
supported by: (a) the maximum available energy, $Q$, for the nine
possible decays is $M(\chi_{b2}^{\prime}) - M(\chi_{b0}) - 2M(\pi^{\pm})$
= 130 MeV, making it difficult to have the extra kinetic energy 
associated with two units of angular momentum; (b) previous studies
of $\Uths$$\to\pi\pi\Uons$\cite{CLEO3Sto1S94} and 
$\psi^{\prime}\to\pi\pi J/\psi $\cite{BES}, systems with
substantially more $Q$, indicate no angular momentum 
between the final state onium and the dipion system, although
the former result is also consistent\cite{QWG} with a few percent 
of $D$-wave; and (c), the average
(weighted by the observed distribution
of di-pion invariant mass, $m_{\pi\pi}$)
of the $D$-wave between the two
pions in $\psi^{\prime}\to \pi\pi J/\psi$\cite{BES} is less than 10\%.

As shown in Table~\ref{tab:allowed}, the entry and exit 
branching fractions\cite{PDG}
strongly disfavor our observation of $J^{\prime} = J = 0$.  
We also had to discriminate against this possible mode in order
to suppress our dominant background source,
``$\pi\pi~\gamma~\gamma$'',
in that there is overlap in the energies of the 
E1 transition photon for the $J^{\prime} = J = 0$ signal
process and that of the dominant $J = 2$ 
mode of that background.  Therefore, we 
assumed that the transitions with $J^{\prime} = J = 1$ or $2$ dominate.
To estimate the relative abundance of these two transitions and, later,
to calculate the partial width $\Gamma_{\pi\pi}$, we needed the {\it full}
widths $\Gamma(\chi_{b2}^{\prime})$ and $\Gamma(\chi_{b1}^{\prime})$.  We
calculated these using the theoretical E1 partial widths for these two
states\cite{KR,EFG} and their experimental 
E1 branching fractions\cite{PDG, CLEOII, CUSB}
%for these two states 
to $\gamma\Upsilon$(1S) and $\gamma\Upsilon$(2S), where in the latter
we took into account the new CLEO III value\cite{Danko} of 
${\cal B}(\Utws\to\mu^{+}\mu^{-})$.  Our results, also listed
in Table~\ref{tab:allowed},
are 
$\Gamma(\chi_{b2}^{\prime}) = (138 \pm 19)$ keV
and 
$\Gamma(\chi_{b1}^{\prime}) = (96 \pm 16)$ keV.
Given 
${\cal B}_{\gamma_{1},J^{\prime}}\cdot{\cal B}_{J,\gamma_{2}}/\Gamma$ 
from this table
we expected the $J^{\prime} = J = 1$ transition to dominate
$J^{\prime} = J = 2$ by roughly a factor of 2.3.

\begin{table}[tbh]
%\begin{ruledtabular}
 \begin{center}
 \begin{tabular}{|c|c||c|c|c|c|c|} 
\hline
$J^{\prime}$ & $J$  & $\Delta$ M & 
${\cal B}_{\gamma_{1},J^{\prime}}$  &
${\cal B}_{J,\gamma_{2}}$ &
$\Gamma$ &
${\cal B}_{\gamma_{1},J^{\prime}}
\cdot{\cal B}_{J,\gamma_{2}} /\Gamma$ \\
(2$P$) & (1$P$) & (MeV) & (\%) & (\%) & (keV) & ($\times 10^{-4}$ keV$^{-1}$) \\
\hline
2& 2 & 356 & 11.4 & 22 & 138 & 1.8 \\
1& 1 & 363 & 11.3 & 35 & 96  & 4.1\\
0& 0 & 372 & 5.4  & $<6$ & -- & --\\
\hline
\end{tabular}
%\end{ruledtabular}
\caption{The three di-pion transitions between $\chi_{b}^{\prime}$ and
$\chi_{b}$ that leave the orbital angular momentum
unchanged ($S$-wave).  The third column is 
the mass difference. Columns four and five are the
branching fractions for the entrance and exit E1 transitions:
${\cal B}_{\gamma_{1},J^{\prime}} = 
{\cal B}(\Upsilon$(3S)$\to\gamma_{1} \chi_{b}^{\prime})$  and 
${\cal B}_{J,\gamma_{2}} =
{\cal B}(\chi_{b}\to\gamma_{2}\Upsilon$(1S)).
The E1 transition from $\chi_{b0}$ is
unobserved, with a limit of 6\% on its branching fraction
at 90\% C.L..}
\label{tab:allowed}
 \end{center}
\end{table}
  
Two approaches were taken to evaluate
$\bothsignal$.
In the first, the ``two-pion'' analysis, we required all the
particles to be found but made minimal requirements on $\gamma_{2}$.
%the photon in $\lowerdecay$.  
A two-dimensional analysis was performed
%with the variables being the 
using the
energy of the photon in
$\upperdecay$, denoted $E_{1}$, and the mass recoiling against
the pion pair, $M_{\mathrm{rec}}$,
to define our signal.
  In calculating $M_{\mathrm{rec}}$ we also
used the four
vector of $\gamma_1$ so that 
%this variable 
$M_{\mathrm{rec}}$
actually 
represents the
mass difference of the $2P$ and $1P$ states; {\it i.e.,}

\begin{equation}
M_{\mathrm{rec}} \equiv
\sqrt{({\cal P}_{3S} - {\cal P}_{\gamma 1})^{2}} - 
\sqrt{({\cal P}_{3S} - {\cal P}_{\gamma 1} - {\cal P}_{\pi 1} -
{\cal P}_{\pi 2})^{2}}~,
\label{eqn:mrecdef}
\end{equation}
with ${\cal P}$ denoting the four-vector momentum.
In the second, we
increased our efficiency by only reconstructing one of the pions
(a ``one-pion'' analysis) and used as variables the missing
mass of the event and $E_1$.

\section{The Channel $\altsignal$}
\label{sec:pippim}
In event selection for our 
study of 
$\altsignal$
we required 
%the proper number of 
four well-measured primary
charged
tracks, two of which had to have high momenta (in excess of 3.75 GeV/c)
and had to have calorimeter and momentum information consistent
with being either $e^{+}e^{-}$ or 
$\mu^{+}\mu^{-}$.\footnote{More 
details on the charged pion analyses are 
available in the MS thesis of K.~M.~Weaver, 
{\it Observation of~}$\altsignal$, 
Cornell University, 2005 (unpublished).} 
These two putative lepton tracks also had
to have an invariant mass within 300 MeV of the $\Uons$ mass, which
is a very loose requirement ($\sim \pm 5\sigma$).
The other track(s)
had to have measured momentum 50 $< p < $ 750 MeV/c
%All tracks were required to have 
%have a reasonable helix fit, come from the luminous volume 
%(radius 3 cm and length 18 cm), 
and
have a dip angle with respect to the
beam axis corresponding to
$|\cos \theta | < 0.93$.
%and have been evaluated to not be a fragment
%of a curling track.  
To reduce QED backgrounds and facilitate
comparison to other, established channels, we made additional, highly
efficient
requirements on the difference of the momenta of the two lepton
candidates and on the maximum allowed momentum of the charged pion
candidates(s).

Transition photon candidates in our analyses of 
$\altsignal$
%, which come from the two E1 transitions,
were defined as calorimeter energy depositions,
in excess of 60 MeV, with lateral profile consistent with
that of a photon, not associated with any charged track or any
known ``noisy'' crystals, and not located in the inner-most portion
of the endcap, roughly 
%defined 
bounded
by $|\cos \theta | \approx 0.93$.  For the
$e^{+}e^{-}$ channel, we further suppressed fragments of the
electron showers.  

\begin{figure}[htbp]
\centerline{\includegraphics{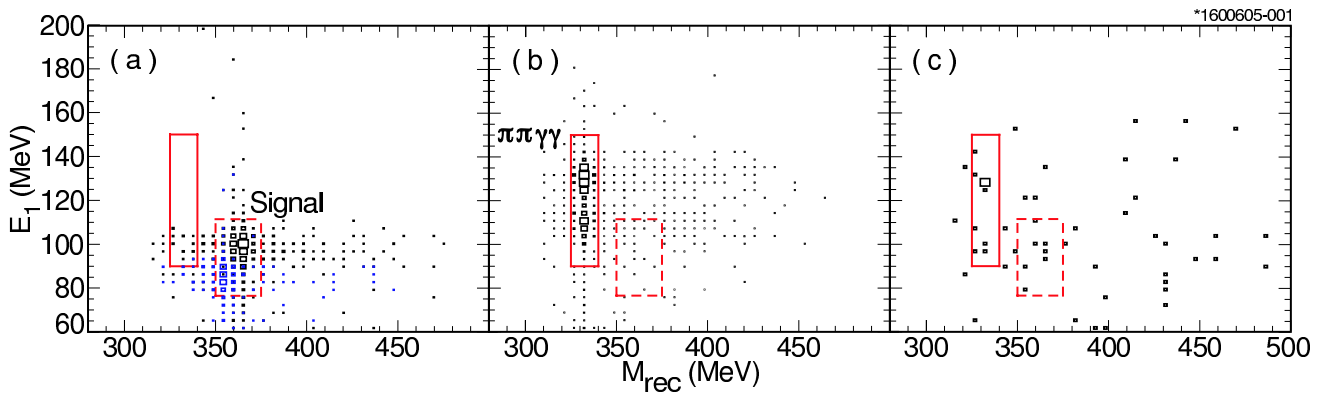}}
\caption{Definition of the three regions in the {\it charged
two-pion} analysis in
the $E_{1}~vs.~M_{\mathrm{rec}}$ plane.
The outline of the figure defines the ``sideband'' region, 
which does {\it not} include the two smaller regions, namely the
solid rectangle in which the $\pi\pi~\gamma~\gamma$ process dominates
and the dashed rectangle 
%as the region for the signal.
in which the signal dominates.
On the left (a) we show Monte Carlo events from both the $1 \to 1$ 
and $2 \to 2$ transitions, with the production ratio of 2.3:1, as
described in the text.  In the middle (b) 
we similarly show Monte Carlo
events from the $\pi\pi~\gamma~\gamma$ process. The data for
this two-pion analysis is shown on the right in (c). }
\label{fig:DipionRegions} 
\end{figure}

In the {\it charged two-pion} (fully-reconstructed) analysis, we required that 
there be 
either 
two or three photon candidates.  If there were two, the higher of the two energies 
had to be in excess of 300 MeV; otherwise $E_{1}$ was deemed
likely to be due to a
spurious calorimeter energy deposition.  
If three were found, then the highest
energy had to exceed 300 MeV and the second highest exceed 120 MeV,
so that it not be confused with a valid $E_{1}$ photon.  Then, based on Monte
Carlo studies of $S^{2}/B$, we defined the three regions shown in
Fig.~\ref{fig:DipionRegions}: a signal region, a region in which
we expect the ``$\pi\pi~\gamma~\gamma$'' process to dominate, and a 
larger ``sideband'' region.  The figure also shows how the
Monte Carlo simulations of signal (left plot) and ``$\pi\pi~\gamma~\gamma$''
(center plot) populate these three regions.
The overall efficiency for the signal
is 5.1\% and 4.3\%, for 
$J^{\prime} = J = 1$ and $J^{\prime} = J = 2$, 
respectively, with the largest
inefficiency coming from 
%having to find 
reconstructing
two high-quality low-momentum 
tracks.  As described in Sect.~\ref{sec:gamma}, these have
relative uncertainties of $\sim$10\%.
% RSG 16Sep05  
% Added sentence about relative ee and mumu efficiencies.
The efficiencies for the $\mu^{+}\mu^{-}$ final state are
10\% (relative) higher than those for the $e^{+}e^{-}$ state in this
analysis; this trend is true for the other three analyses in this
Article as well.
% RSG 16Sep05 

\begin{table}[htb]
\begin{center}
\begin{tabular}{|c||c|c|c||c|c|c|}
\hline 
Region of & Estimated & Constrained & Number & 
Estimated & Constrained & Number \\
Plot & Occupancy & Occupancy & Observed &
Occupancy & Occupancy & Observed\\
\hline
~~ 
%& \multicolumn{3}{|c||}{$\pi^{+}$ and $\pi^{-}$ found} 
& \multicolumn{3}{|c||}{$\pi^{+}\pi^{-}$ found} 
& \multicolumn{3}{|c|}{$\piz\piz$ found}\\
\hline
Sideband                    
& $22.7 \pm 4.4$ & $36$           & 36 
& $16.5 \pm 2.4$ & $15$           & 15  \\
$\pi\pi~\gamma~\gamma$   
& $ 8.6 \pm 2.0$ & $ 9.0 \pm 2.0$ & 10 
& $ 13.7 \pm 3.2$ & $ 13.6 \pm 3.2$ & 15  \\
Signal                   
& $ 0.6 \pm 0.2$ & $ 1.0 \pm 0.3$ &  7  
& $ 2.3 \pm 0.5$ & $ 2.2 \pm 0.5$ &  1 
\\
\hline
~~ 
& \multicolumn{3}{|c||}{One $\pi^{\pm}$ found} 
& \multicolumn{3}{|c|}{One $\piz$ found} \\
\hline
Sideband                    
& $ 5.2 \pm 1.4$ & $8$           & 8
& $ 15.2 \pm 3.4$ & $17$           & 17 \\
$\pi\pi~\gamma~\gamma$  
& $17.1 \pm 4.8$ & $18.0 \pm 4.9$ & 26  
& $14.4 \pm 3.2$ & $14.8 \pm 3.4$ & 13  \\
Signal
& $ 2.2 \pm 0.6$ & $ 2.6 \pm 0.7$ & 17 
& $ 26.5 \pm 5.7$ & $ 26.9 \pm 5.8$ & 35  \\
\hline
\end{tabular}
\caption{The results of the four analyses, showing the
predicted occupancies in each of the three regions of
interest and the observed number of events in those
regions.  In the ``constrained'' column the predictions have
been adjusted to make that of the sideband region match the
observed number in that region.}
\label{tab:results}
\end{center}
\end{table}
We also show in the same figure the {\it data} for this two-pion analysis,
which has 36/10/7 events in the 
sideband/$\pi\pi~\gamma~\gamma$/signal regions,
respectively.  

Using Monte Carlo simulations of the $\Uths$ decays and the
high-energy continuum data, all properly scaled, we predict the
number of expected events in the three regions, as shown in 
Table~\ref{tab:results}; the uncertainties listed are from the
various branching fractions\cite{PDG} used in the scaling to our
accumulated number of $\Uths$ decays.  The prediction for the
``$\pi\pi~\gamma~\gamma$'' region (the second line of the table)
is very consistent with the observation in 
data, which also has roughly equal numbers of $e^{+}e^{-}$ (4) and
$\mu^{+}\mu^{-}$ (6) final states.
The large sideband region prediction is somewhat smaller than
the data, particularly in the $e^{+}e^{-}$ final state.  
To take a more conservative
approach to the number of events 
expected in the signal region due to known
processes and backgrounds, we then added in enough events, scaled in
proportion to the size of each box, to bring the sideband region into
exact balance.\footnote{For example, for the charged two-pion analysis,
which is in the upper left portion of Table~\ref{tab:results},
the {\it excess} in the sideband region is 36 (observed) minus 22.7
(estimated) or 13.3 events.  Scaled by the relative areas, this 13.3
increment means
an additional 0.4 events from this potential background source
for each of the two smaller regions.}  
This procedure is labeled ``constrained occupancy''
in Table~\ref{tab:results}; it predicts $1.0 \pm 0.3$ events in the 
signal region, in which we observe
seven, of which six are $\mu^{+}\mu^{-}$.  
% RSG 16Sep05
% This section eliminated in that it is redundant with text in Section IV.A
%To estimate the
%significance of the excess we used a simple  Monte Carlo 
%procedure with mean background
%levels normally distributed about 1.0 with standard deviation of 0.3 and
%determined the probabilities of observing seven or more events.  The
%null hypothesis, namely that backgrounds can account for the observation,
%has a probability of $2.2 \cdot 10^{-4}$.
% RSG 16Sep05

In addition to observing that 
%we properly populate 
the $\pi\pi~\gamma~\gamma$ region is properly populated
($8.6 \pm 2.0$ events expected {\it vs.} 10 observed), 
we checked that our analysis procedures, when instead requiring
0 or 1 photon, can reproduce the measured product branching fraction
${\cal B}(\Uths\to\pi^{+}\pi^{-}\Utws)$ 
$\cdot$
${\cal B}(\Utws\to\ell^{+}\ell^{-})$, which is,
by weighted average of the results of CLEO~I\cite{Brock},
CLEO~II\cite{CLEO3Sto1S94} and CUSB\cite{Wu},
$(1.10\pm 0.12)\cdot 10^{-3}$.  We observed
$154 \pm 13$  $\mu^{+}\mu^{-}$ such events and $152 \pm 39$  
$e^{+}e^{-}$, which implies an efficiency in the CLEO III 
{\it data} of $(4.8 \pm 0.8)$\%.
Our Monte Carlo simulations of this channel indicate an efficiency of
$(4.3 \pm 0.1)$\%, in agreement with the data.

Given the low efficiency for finding low momentum pions, our second
approach (the {\it charged one-pion} analysis) 
was to require only one 
%such 
soft charged track but make tighter 
demands on $\gamma_{2}$ (see Fig.~\ref{fig:signal_and_TS}) and on the
lepton pair.  The sum of the measured $E_{1}$ and $E_{2}$ was fit to 
%be 
518 MeV, the properly weighted average for that sum from our Monte
Carlo simulation of the signal, yielding a $\chi^{2}_{\gamma\gamma}$ 
for the fit as 
a figure of merit.  We required $\chi^{2}_{\gamma\gamma} < 4$.  The 
momenta of the lepton
pair were used in a mass-constrained fit to the $\Uons$ mass, for
which we required $\chi^{2} < 10$.  In constructing the missing mass
of the event, which for signal would be $M(\pi)$, we used
as inputs 
the $\Uths$ mass, 
the angles and fitted energies of the two photons, 
${\cal P}_{\Upsilon}$ (the momentum four vector of the
fitted $\Uons$),
and the momentum of the one measured charged pion:

\begin{equation}
M_{\mathrm{miss}} \equiv
\sqrt{({\cal P}_{3S} - {\cal P}^{\mathrm{fit}}_{\gamma 1}
- {\cal P}^{\mathrm{fit}}_{\gamma 2}
- {\cal P}_{\Upsilon}
- {\cal P}_{\pi})^{2}}~.
\label{eqn:mmissdef}
\end{equation}

\begin{figure}[htb]
\centerline{\includegraphics[width=10.0cm]{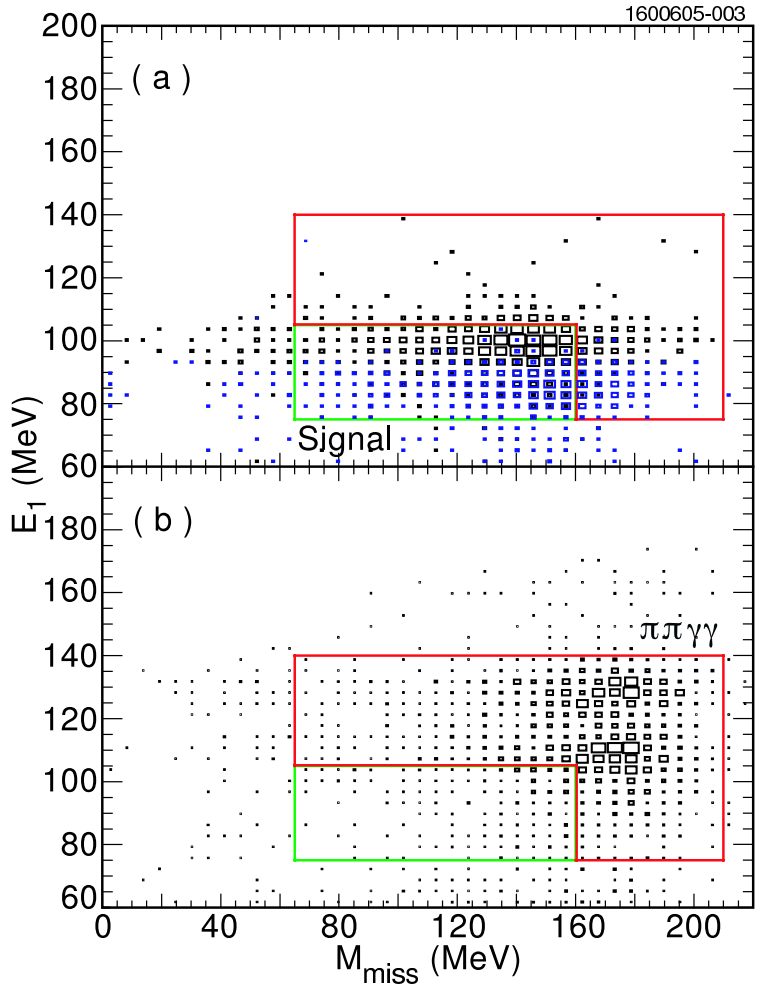}}
\caption{Definition of the three regions in the {\it charged 
one-pion} analysis in
the $E_{1}~vs.~M_{\mathrm{miss}}$ plane.
In both cases the outline of the figure is the large ``sideband'' region.
On the top (a) we show the smaller ``signal'' region and the 
``$\pi\pi~\gamma~\gamma$'' region,
and the Monte Carlo events from both the $1 \to 1$ 
and $2 \to 2$ transitions.  On the 
bottom (b) we similarly show both of these regions and the Monte Carlo
events from the 
%two photon cascade 
``$\pi\pi~\gamma~\gamma$'' process.}
\label{fig:SinglePionRegions} 
\end{figure}

Given that we only observed one of the pions, the calculated di-pion
invariant mass for the charged one-pion analysis, namely,
%, $m_{\pi\pi}$, 

\begin{equation}
m^{2}_{\pi(\pi)} = ({\cal P}_{\pi}+ {\cal P}_{\mathrm{miss}})^{2}
= ({\cal P}_{3S}- {\cal P}_{\Upsilon} - 
{\cal P}^{\mathrm{fit}}_{\gamma 1} -
{\cal P}^{\mathrm{fit}}_{\gamma 2})^{2}~,
\label{eqn:mpipi1}
\end{equation}
% Changed _bm to _3S to match other equations - RSGalik - 04Jan06
was not constrained to be in excess
of twice the pion mass.  Simulations show a selection criterion
of $m_{\pi(\pi)} > 260$ MeV to be highly efficient for 
%the data signal 
$\altsignal$, and this was applied to minimize backgrounds.

We again used a study of $S^{2}/B$ to determine a signal region,
this time in the $E_{1}~vs.~M_{\mathrm{miss}}$ plane, as depicted in 
Fig.~\ref{fig:SinglePionRegions}. The region assigned to the main background,
``$\pi\pi~\gamma~\gamma$'', was somewhat larger and contiguous to the
signal region;
the boundaries were selected 
to have the sideband region have as
few events as possible that were either signal or this main background.
We found from our Monte Carlo simulations that the overall
efficiency for this one-pion analysis is 
10.6\%
for
$J^{\prime} = J = 1$ and
9.6\%
for $J^{\prime} = J = 2$.
% efficiencies checked against p.15 of CBX05-46 and ToyMC_charged.C
As detailed in Sect.~\ref{sec:gamma} the 
(relative)
uncertainties in these
efficiencies are roughly 10\% and 8\%, respectively.

\begin{figure}[tbh]
  \centerline{\hbox{     
       \includegraphics{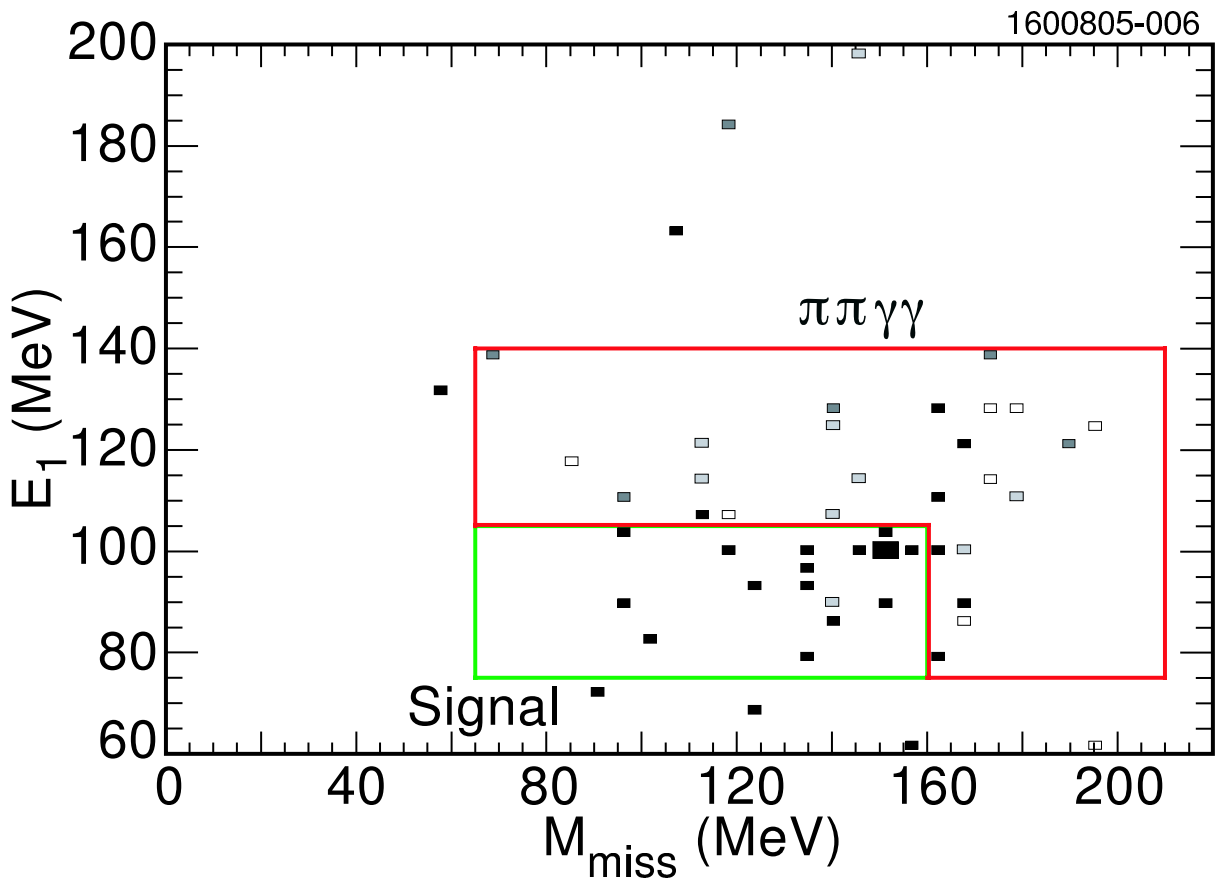}
    } % close \hbox{
   }  % close \centerline{
 \caption{The data 
events falling into our three defined regions for the 
{\it charged one-pion} 
analysis, shaded according to their $\chigg$ value.  
The darkest boxes are for events with $\chigg$ values between 0 and 1 and the
lightest for those with $\chigg$ values between 3 and 4.
%Black markers indicate events
%with $\chigg$ values between 0 and 1, light grey between 1 and 2, solid lines between 2 and 3, 
%and dotted
%lines indicate events with $\chigg$ values between 3 and 4.  
Most of the events in the small signal 
region (the smaller rectangle)
show low values for $\chigg$, indicating excellent fits of the photon
energy sum to that expected for signal events.}
 \label{fig:comparison} 
\end{figure}

The data are shown in Fig.~\ref{fig:comparison} and the yields are
listed in Table~\ref{tab:results}.  Of the 17 signal events, nine
have $\ell = \mu$ and the other eight have $\ell = e$.  The 
population of the ``$\pi\pi~\gamma\gamma$'' region is consistent with, 
although a bit larger than, our prediction.

The sideband region also has a somewhat larger yield than predicted, so,
as in the
two-pion analysis we added in enough background events to balance the
sideband region, as shown in the ``constrained'' column of 
Table~\ref{tab:results}.
The probability that the backgrounds, constrained to give the
sideband yield, could 
produce the observed population in the signal region
is $1.3\cdot10^{-7}$.  

The distribution of $\chi^{2}_{\gamma\gamma}$ for the 17 events in the signal
region closely mimics that seen in our Monte Carlo simulation.  The
values of $\chi^{2}_{\gamma\gamma}$ are encoded in Fig.~\ref{fig:comparison},
showing a predominance of low ({\it i.e.,} better) 
values of this figure of merit for the
events in the signal region. To further test this aspect of the analysis
we 
%again ran over the data, 
instead optimized our selection criteria and our
constraints for the ``$\pi\pi~\gamma~\gamma$'' process.  We found 49 events,
which imply an efficiency in data of $(15.0\pm 4.8)\%$; our Monte Carlo
simulations predicted an efficiency for this test of $(12.7\pm 0.2)\%$,
showing consistency. 

We also checked that there is not some other, resonance-induced effect
that could mimic our charged 
one-pion signal by 
analyzing $\Utws$ data, 
with
selection criteria and plotted variables appropriately scaled to account
for the mass difference between the $\Utws$ and the $\Uths$
resonances. Only
three events passed our selection criteria with none of them in the
signal region; we assumed no such background sources in further
analysis.

\section{The Channel $\neutsignal$}
\label{sec:pizpiz}

Most of the selection criteria for our study involving neutral pions
were the same as those in the preceding section.  Given the relatively
small $Q$ of our process, the photons from the 
transition
$\piz$ decays
tend to be of low energy.  Therefore, we lowered the energy cut-off in the
calorimeter barrel ($|\cos\theta | < 0.80$) to 30 MeV. 
% RSG 16Sep05
% Reworded as per CLEOAC
In the endcap
regions ($0.80 < |\cos\theta | < 0.93$) 
photons were still required to have energy in excess
of 60 MeV.
% RSG 16Sep05
In addition to these photon candidates, we also allowed one
endcap shower in the energy range
$30 < E < 60$ MeV to be used as a decay product
of the neutral pions. 
All $\piz$ candidates were 
formed from high quality showers that were not associated
with charged tracks, with the exception that no
$\piz$ candidate could
%found by standard CLEO software and cannot 
use the highest energy photon in the event, which
was presumed to be from the transition $\chi_{b}\to\gamma\Upsilon$(1S).

For the fully reconstructed, {\it neutral two-pion} analysis, 
there had to be five or more photon candidates and no charged tracks 
other than the two lepton candidates.
%and .  
%There had to be 
We require
two or more found $\piz$
candidates, of which we kept the best two based on their goodness
of fit to the $\piz$ mass hypothesis 
($S_{\pi^{0}}\equiv (M_{\gamma\gamma} - M_{\piz})/\sigma$).  
We further ensured
good $\piz$ candidates by requiring that the sum of the squares of the 
two pulls ({\it i.e.,} the two
$\chi^{2}$ values from the $\pi^{0}$ fits) be less
than 25.
All photon candidates not used in forming the two neutral pions
were then investigated in pairs in a fit to 518 MeV,
the
expected sum of the transition energies $E_{1}$ and $E_{2}$.  
The best pair was kept; the chi-square of the fit was
restricted to $\chi^{2}_{\gamma\gamma} < 9$. 

%Studies of the 
Our
simulations indicate that the three regions used 
in the {\it charged} two-pion analysis were also optimal for the
neutral case, with distributions of the signal and primary 
background similar to those in Fig.~\ref{fig:DipionRegions}.  
We found 
%Based on the Monte Carlo simulations 
the efficiency for a $\pi^{0}\pi^{0}$
signal with $J^{\prime} = J = 1$ is 7.2\% and for
$J^{\prime} = J = 2$ is 6.4\%, with roughly 11\% (relative)
uncertainties. 

Again using the known branching fractions\cite{PDG}, we can
predict the occupancies of these three regions in the absence of
a signal, as shown in the $\piz\piz$ section of Table~\ref{tab:results}.
While the sideband and ``$\pi\pi~\gamma~\gamma$'' regions have the expected
populations,
there is only one event in the signal region.  This analysis
supports the null hypothesis, with roughly a 90\% probability that
the predicted occupancy of 2.3 events would give one or more
events in that region.  

%Last, we investigated 
For
the {\it neutral one-pion} analysis  
the fit of the $\piz$  candidate 
%$\piz$ 
had to be 
in the range $-7 < S_{\piz} < 7$. 
Because in a typical event there are several photons of energy near 100 MeV
and because we required {\it exactly} one found $\piz$, a large combinatoric
``$\pi\pi~\gamma~\gamma$'' background can 
contaminate
%feed into
the signal region.  

The highest energy photon in the event was required to have $E > 370$ MeV.
It was
then paired 
with all other photon candidates not used in forming the 
lone $\piz$ to find the best match to the photon energy
sum of 518 MeV; for this
neutral analysis we require 
$\chi^{2}_{\gamma\gamma} < 3$ (the limit was 4 in the corresponding 
charged
analysis).

Requirements on the reconstructed pion and on the lepton candidates
were similar to those of the charged one-pion analysis.  In addition
we required that the energy of the missing $\piz$, based on the energies of the
found particles, be in the range $100 < E_{\mathrm{miss}} < 240$ MeV.

The regions for the signal and primary backgrounds from the
charged one-pion 
analysis were {\it not} found to be optimal in the neutral
case.  While the sideband region remained the same, $S^{2}/B$ studies
showed the optimal signal region to have $75 < E_{1} < 110$ MeV
and $65 < M_{\mathrm{miss}} < 210$ MeV and the ``$\pi\pi~\gamma~\gamma$''
region to best be $110 < E_{1} < 140$ MeV and, again, 
$65 < M_{\mathrm{miss}} < 210$ MeV.  For these selection criteria the efficiencies
are 13.4\% for $J^{\prime} = J = 1$ and 12.3\% for $J^{\prime} = J = 2$;
the relative uncertainties are roughly 16\% and 12\%,  respectively.

The results from the data are shown in Table~\ref{tab:results}.
We find the occupancies of the two non-signal regions again to be 
near our expectations.  For the signal region there is a slight excess, 
with 35 events being observed but only $26.9 \pm 5.8$ expected. 

\section{Discussion of Results and Summary}
\label{sec:resultsandsummary}

\subsection{The Null Hypothesis}
\label{sec:null}

%Discussion of the null hypothesis.
We now have four analyses on which to base our test of the
null hypothesis that the backgrounds alone can account for the
observed data in the signal regions.  
%One must be careful to not
%simply multiply an ever increasing  
%number of probabilities to get a smaller and smaller 
%overall value without recognizing that each such probability, if the
%null hypothesis were correct, would average at 0.50. 

In all cases we use the predicted occupancies (which represents
the null hypothesis) from the ``constrained'' column of
Table~\ref{tab:results},
thus allowing for the fact that there may be some background contribution
%for which we did not account with 
unaccounted for by
our simulations and continuum data
samples.  From this table we generate a large
number of experimental mean occupancies and use Poisson statistics
to assess the 
statistical consistency
%probability 
of backgrounds 
alone with
%accounting for
the number of observed events in data (or more).  
%Each
%entry below represents {\bf one} test; these can be combined in various
%ways for publication purposes.

%\begin{itemize}

%\item $\pi^{+}\pi^{-}$ \\
{\it Charged two-pion analysis:}
For example, here we create many experiments that have a 
Gaussian-distributed background
level with mean of 1.0 events and standard
deviation of 0.3 events. 
%and determine t
The Poisson probability for this
to result in 7 or more observed events is 
%This is from Tables 15 and 18 in CBX05-19
%and also represented in Table~\ref{tab:combined_results_dipion}.
%The probability for $1.0 \pm 0.3$ events to yield 7 or more is
$2.2 \cdot 10^{-4}$, or a one-sided Gaussian effect at 3.5$\sigma$.

%\item $\pi^{\pm}$ \\
{\it Charged one-pion analysis:}
The probability for $2.6 \pm 0.7$ events to yield 17 or more is
$1.3 \cdot 10^{-7}$, or a one-sided Gaussian effect at 5.2$\sigma$.

%\item $\piz\piz$ \\

{\it Neutral two-pion analysis:}
Similarly, $2.2 \pm 0.5$ events have a 87\% probability of 
accounting for the lone signal event, thus supporting the null
hypothesis.
% This had said 2.5 /pm 0.5 (?) - RSGalik - 04Jan06
%
%As seen in Table~\ref{tab:combined_results_dipion} we expect
%about 2.3 events of occupancy from background sources but observe
%only one event in the signal region.  As is ingrained in most of
%us, a mean of 2.3 events has a 10\% pro-ability of yielding an
%observation of zero.  Thus, in terms of the null hypothesis,
%there is a 90\% probability of the backgrounds accounting for
%the observation of one or more events.

%\item $\pi^{0}$ \\
{\it Neutral one-pion analysis:}
%For this small excess we use Gaussian statistics that show
%a probability of  $8 \%$ for the
%null hypothesis.
The null hypothesis has an 8\% probability of accounting for
the yield in this analysis.

%\end{itemize}

One can combine the charged and neutral  two-pion analyses into
{\it one} test: they are statistically independent
and have the same signal region contour.
Summing the entries in Table~\ref{tab:results}, the
probability for $3.3 \pm 0.6$ events to yield 8 or more is
$2.6 \%$.  
% RSG 16Sep05
% Added a statement about the isospin situation
% removed the original reminder about the one-pion analyses
%Coupled with the 8\% probability from the neutral one-pion
%analysis and the $1.3 \cdot 10^{-7}$ probability from the
%charged one-pion analysis, we conclude that the null
%hypothesis is not substantiated.

The analyses with charged pions show a pronounced signal that is 
supported by the neutral one-pion analysis.  Given our predicted 
backgrounds and the partial width inferred from the charged pion 
analyses, there is a 2\% probability of only
seeing zero or one event in the neutral two-pion study.  
Taking all 
four analyses together, we conclude that the null hypothesis is not 
substantiated.
% RSG 16Sep05

%  This sub-section removed 01Sep05 after committee meeting - RSGalik
%\subsection{The Di-pion Invariant Mass}
%\label{sec:mpipi}
%
%Given that the two analyses involving neutral pions do not give
%meaningful excesses over background, we use only the two charged
%analyses to evaluate the di-pion invariant mass, $m_{\pi\pi}$.
%In Fig.~\ref{fig:mpipi} we show this distribution
%%$m_{\pi\pi}$,
%for the $7+17=24$ signal events of the combined charged pion 
%analyses.  For the two-pion
%approach this uses the measured pion momenta; for the one-pion analysis
%$m_{\pi\pi}$ is defined as:
%
%\begin{equation}
%m^{2}_{\pi(\pi)} = ({\cal P}_{\pi}+ {\cal P}_{\mathrm{miss}})^{2}
%= ({\cal P}_{bm}- {\cal P}_{\Upsilon} - 
%{\cal P}^{\mathrm{fit}}_{\gamma 1} -
%{\cal P}^{\mathrm{fit}}_{\gamma 2})^{2}.
%\label{eqn:mpipi1}
%\end{equation}
%
%\begin{figure}[htbp]
%\centerline{\hbox{ \hspace{0.2cm}
%\includegraphics[width=14cm]{DiMass_Separate.eps}
%} % close \hbox{
%}  % close \centerline{
% \caption{The distribution of the di-pion invariant mass,
%$m_{\pi\pi}$ for data in the signal region.  We show this separately for the 
%two-pion analysis (lighter) and the one-pion analysis (darker).  
%The vertical dotted lines
%delineate the physical region; the lower limit is $2m_{\pi}$ for
%both cases; the upper limit depends on the $J$ value.}
% \label{fig:mpipi} 
%\end{figure}
%
%Given that there are only 24 events of which four are likely background
%and that the $m_{\pi\pi}$ values are 
%significantly smeared for the one-pion analysis we
%do not attempt to fit this distribution.
%
% End os deleted sib-section
\subsection{Partial Width for the Di-pion Decay}
\label{sec:gamma}

%Discussion of the partial width (still left form the old draft).
%Much work to do here!!!

\begin{table}[hbt]
\begin{center}
\begin{tabular}{|l||c|c|c||c|c|c|}
\hline 
Uncertainty
&\multicolumn{3}{|c||}{Charged}
&\multicolumn{3}{|c|}{Neutral} \\
\cline{2-7}
Source & 
$\Delta\epsilon /\epsilon$ (\%) & 
$\Delta \epsilon_{1\to 1}$ &
$\Delta \epsilon_{2\to 2}$ &
$\Delta\epsilon /\epsilon$ (\%) & 
$\Delta \epsilon_{1\to 1}$ &
$\Delta \epsilon_{2\to 2}$ \\
\hline
Limited MC statistics              
& -   & 0.3  & 0.3  
& -   & 0.3  & 0.3  \\
Running period dependence         
& -   & 0.5   &  small  
& -   & small  &  small \\
%Position of $E_1$ peak                 
%& -   & 0.5   &  small  
%& -   & -    &  -   \\
Signal region definition           
& -   & 0.6   & 0.1   
& -   & 1.7   & 1.0   \\
Shape of $m_{\pi\pi}$ distribution 
& 2  & -    & -    
& 1  & -    & -    \\
Decay angular distribution           
& 2   & -    & -    
& 2   & -    & -    \\
$\piz$, $\pi^{\pm}$ and $\ell$ finding             
& 6   & -    & -    
& 8   & -    & -    \\
Photon-finding probability             
& 2   & -    & -    
& 2   & -    & -    \\
%Effects of soft end-cap photons
%& -  & - & -
%& N/A  & - & -\\
$\ell = e/\mu$ selection                   
& 1 & - & - 
& 1 & - & - \\
Other selection criteria           
& -   & small & small 
& -   & small & small \\
\hline  
Sum                                
& 7\% & 0.8 & 0.3 
& 10\% & 1.7 & 1.0 \\
\hline
\end{tabular}
\caption{The systematic uncertainties in the efficiencies
for the two one-pion analyses.  For the correlated efficiencies
these are listed as a relative percentage; for the individual
uncorrelated effects, absolute values are shown.}
\label{tab:systeff}
\end{center}
\end{table}

Assuming that our data constitute observation of the
signal process, we 
%can 
then proceed to obtain values 
for the partial
width for this di-pion transition.
We assume there are no $D$-wave contributions and
that our observation of the $J^{\prime} = J = 0$ transition is 
suppressed (see Table~\ref{tab:allowed} and the associated
discussion).  
Here we use\cite{KY1981}
$\Gamma_{\pi\pi} \equiv
\Gamma(\chi_{b1}^{\prime}\to\pi\pi\chi_{b1}) =
\Gamma(\chi_{b2}^{\prime}\to\pi\pi\chi_{b2})$.
Invoking isospin as a good quantum number in
such strong interaction decays, and neglecting the 
small effects of the $\pi^{\pm} - \piz$ mass
difference, we also have
$\Gamma_{\pi\pi} =  \frac{3}{2}\cdot\Gamma_{\pi^{+}\pi^{-}}
= 3\cdot\Gamma_{\pi^{0}\pi^{0}}$. 
%(both
%the E1 branching fractions shown in Table~\ref{tab:allowed} and the
%selection criteria necessary to differentiate signal from the
%$\pi\pi~\gamma~\gamma$ background)
We then write:

\begin{equation}
N_{\mathrm{sig}} = N_{\Upsilon (3S)} 
\cdot {\cal B}_{\Upsilon\to\ell^{+}\ell^{-}}
\cdot \frac{C}{3}\Gamma_{\pi\pi}
\cdot \left[
\frac
{{\cal B}_{\gamma 1,1}~\epsilon_{1\to 1}~{\cal B}_{1,\gamma 2}}
{\Gamma(\chi_{b1}^{\prime})} +
\frac
{{\cal B}_{\gamma 1,2}~\epsilon_{2\to 2}~{\cal B}_{2,\gamma 2}}
{\Gamma(\chi_{b2}^{\prime})} \right] ,
\label{eqn:defining}
\end{equation}
with $C = 1$ or 2 in the neutral and charged cases, respectively.
Here, $N_{\mathrm{sig}} = 
N_{\mathrm{obs}} - N_{\mathrm{bck}}$; the second term is the weighted average
of the two background estimation schemes in Table~\ref{tab:results},
and the difference of this average from either scheme 
%will be
is
included in the systematic uncertainty.  The only
statistical uncertainty 
%will be 
is in the number of observed events, 
$\sqrt{N_{\mathrm{obs}}}$.
We use 
${\cal B}_{\Upsilon\to\ell^{+}\ell^{-}} = (4.96\pm 0.12)\%$,
and the four E1 transition branching fractions are as in Ref.\cite{PDG}
and Table~\ref{tab:allowed}. 
The uncertainties in these ${\cal B}$ values, in the values of
$\Gamma$ as given earlier,
and in the number of parent $\Uths$ 
%will be 
are
taken as systematic in nature.  These are effectively ``common'' to all
four analyses at the level of 20-24\%, depending slightly on the relative ratios
of the two efficiencies in Eqn.~\ref{eqn:defining}.  
The uncertainties in the level of background to be subtracted and 
in the two efficiencies are ``particular'' to each of the analyses.

The contributions to the systematic
uncertainties to the efficiencies in the two
one-pion analyses are shown in Table~\ref{tab:systeff}, with those for
the two-pion analyses being similar in source and magnitude.  
%Many of the
%contributions are the same for both the $J = 1$ and $J = 2$ transitions,
%while others can be unique to one or the other.
As evident 
from Fig.~\ref{fig:SinglePionRegions},
the selection criterion on $E_{1}$ is very tight for the $J^{\prime} = J = 1$
transition in the charged one-pion case, leading 
to significant uncertainty in the modeling of that process; this
is similarly problematic for the neutral one-pion analysis.  For the
$J^{\prime} = J = 2$ transition, the photons from the unfound $\piz$ in the 
neutral one-pion analysis lead to a sizeable uncertainty in our modeling
near the {\it lower} boundary of the signal box.  

We have varied the di-pion invariant mass distribution in the
Monte Carlo simulations to include
three-body phase space, a Yan distribution\cite{Yan} and
a flat distribution, and found relative efficiency variations of from 1\% 
(for the charged case) to 2\% (for the neutral case).  We have included in our stated
efficiencies the effect of the angular distribution
of the transition photon in $\Uths \to \gamma \chi_{b}^{\prime}$
not being isotropic; this is roughly a 2\% effect.  We have {\it not}
included such effects for the decay $\chi_{b} \to \gamma \Uons$, and 
posit a 2\% uncertainty for this source.  For our ability to model the
detection of the transition photons we assign an additional 
systematic uncertainty of 1\% per photon.

We take a 1\% per track systematic uncertainty for finding the high
momentum leptons\cite{Li}.  For the softer pion tracks we take a
2\% per pion uncertainty; this is substantiated by CLEO studies
of charged di-pion transitions in the charmonium system and our
own checks of the overall efficiency presented in Sect.\ref{sec:pippim}.
Neutral pion finding efficiencies are checked in CLEO studies
of neutral di-pion transitions in the $\Upsilon$ and charmonium
systems, for which we assign 3\% per pion as the systematic
uncertainty for finding (or not finding) a $\piz$.  These particle
finding uncertainties are conservatively added linearly in the Table.

There is a small uncertainty in our ability to model the
lepton identification requirements\cite{Li}, which we conservatively
take as 1\%.
The other entries are
%in Table~\ref{tab:systeff} are either
%standard CLEO estimates or 
found to be negligible,
given the generally loose nature of the selection criteria.

To determine the resultant systematic uncertainty for 
$\Gamma_{\pi\pi}$ we use a toy Monte Carlo, generating all the
inputs in Eqn.~\ref{eqn:defining} distributed as Gaussians 
with their uncertainties, and ask
for the region that symmetrically bounds $68.3 \%$ of the values.

As discussed in Sect.~\ref{sec:null}, we evaluated 
$\Gamma_{\pi\pi}$ for three
situations: charged one-pion,  neutral one-pion, and combined two-pion. 

\begin{table}[hbt]
\begin{center}
\begin{tabular}{|l||c|c|c|c|c|}
\hline 
% RSG 30Aug05: Table remade after redoing calculations; a few small
%    inconsistencies made correct; kept 3 decimal places
Channel & $N_{\mathrm{obs}}$ & $N_{\mathrm{bck}}$ & 
$\epsilon_{1\to 1}$ (\%)& $\epsilon_{2\to 2}$ (\%)& 
$\Gamma_{\pi\pi}$ (keV)\\
\hline 
Charged one-pion & 17 & $2.4 \pm 0.7$ & 
$(1 \pm 0.07)(10.6 \pm 0.8)$& $(1 \pm 0.07)(9.6 \pm 0.3)$ & 
$1.24 \pm 0.35 \pm 0.12$ \\
% RSG 30Aug05: 1.243 \pm0.351 +0.132-0.112
% RSG 02Sep05: added additional syst as per committee; 0.6 -> 0.7
Neutral one-pion & 35 & $26.7 \pm 5.8$ &
$(1 \pm 0.10)(13.4 \pm 1.7)$& $(1 \pm 0.10)(12.3 \pm 1.0)$ &
$1.12 \pm 0.80~^{+0.82}_{-0.78}$ \\
% RSG 30Aug05: 1.115 \pm0.795 +0.805-0.765
% RSG 30Aug05: for second error will use \pm 0.79 when making averages
% RSG 02Sep05: added additional syst as per committee; 5.7 -> 5.8;
%        0.81 -> 0.82; 0.77 -> 0.78; will use \pm 0.80 for averaging
Charged two-pion & -- & -- & 
$(1 \pm 0.10)\cdot 5.1$ & $(1 \pm 0.10)\cdot 4.3$ &
~ \\
Neutral two-pion & --& --&
$(1 \pm 0.11)\cdot 7.2$ & $(1 \pm 0.11)\cdot 6.4$ & 
~ \\
Combined two-pion & 8 & $3.1 \pm 0.6$ &
-- & -- & 
$0.52 \pm 0.30 \pm 0.08$\\
% RSG 30Aug05: 0.516 \pm0.298 \pm0.074
% RSG 02Sep05: added additional syst as per committee; 0.5 -> 0.6;
%          \pm0.074 -> \pm 0.081
\hline
\end{tabular}
\caption{The various contributions to the calculation of the
partial width from sources in this experiment.  The two two-pion
analyses have been combined for the width determination.  Of the 
two quoted uncertainties, the first is statistical and the second
is from the uncertainties in $N_{\mathrm{bck}}$ and the efficiencies.  An
additional systematic uncertainty of $\sim 22\%$ comes from
branching fractions, estimates of the total widths, and the
number of $\Uths$.}
\label{tab:contributions}
\end{center}
\end{table}

The individual contributions to Eqn.~\ref{eqn:defining} are shown in
Table~\ref{tab:contributions}, along with the value of $\Gamma_{\pi\pi}$
obtained, its statistical uncertainty, and its individual (CLEO-based) 
systematic uncertainty.  Taking the {\it statistical} average of the three
gives
$\Gamma_{\pi\pi}{\rm [stat~only]} = (0.84 \pm 0.22)$ keV.  A more
%RSG 30Aug05: 0.843 \pm 0.218
complete average takes into account the individual systematic
uncertainties; this weighted average is
$\Gamma_{\pi\pi}  = (0.83 \pm 0.23)$ keV.  
%RSG 30Aug05: 0.830 \pm 0.232
%RSG 02Sep05: 0.826 \pm 0.233
Following the PDG\cite{PDG} prescription for evaluating the
consistency of results being averaged, we obtain an 
uncertainty scale factor of 1.09, which is close to unity.

De-convolving the 
%0.22
statistical uncertainty and adding in a separate term for 
the 22\% ``common'' uncertainties
that are not based on measurements in this analysis
yields our final result of
$$
%\begin{equation}
%\Gamma_{\pi\pi} {\rm [overall~result]} = 
\Gamma_{\pi\pi} = 
(0.83 \pm 0.22 \pm 0.08 \pm 0.19)~{\rm keV}~,
%RSG 30Aug05: 0.830 \pm 0.218 \pm 0.081 \pm 0.187
%RSG 02Sepo5: 0.826 \pm 0.218 \pm 0.082 \pm 0.186
%\label{eqn:finalresultoverall}
%\end{equation}
$$
with the 
%last 
uncertainties being statistical,
systematics from our analyses, and systematics from
outside sources.
% RSG 16Sep05
% Comparisons added as requested by CLEOAC
This result for $\bothsignal$ can be compared to values derived from 
the PDG\cite{PDG} of
$\Gamma(\Uths\to\pi\pi\Utws) = (1.3 \pm 0.2$) keV for a 
process with somewhat less $Q$ and
$\Gamma(\Utws\to\pi\pi\Uons) = (12 \pm 2)$ keV for a 
process with considerably more $Q$.  
% RSG 16Sep05
% RSG 05Oct05
% Added the numerical results from Table II of the K-Y paper.
Our result is 
consistent with the theoretical expectations of
Kuang and Yan\cite{KY1981}, who have calculated
$\Gamma_{\pi\pi} = 0.4$ keV. 
% RSG 05Oct05

In summary, we have searched the CLEO III data at the $\Uths$
resonance for the decay $\bothsignal$ using four 
different approaches. The
combined probability that the signal process is absent is 
small, leading to the conclusion that the
null hypothesis cannot be substantiated.  Under the assumption of
no $D$-wave contributions we obtain a partial width for each of 
the $J^{\prime} = J = 1$ and $J^{\prime} = J = 2$ transitions of 
$\Gamma_{\pi\pi} = 
(0.83 \pm 0.22 \pm 0.08 \pm 0.19)~{\rm keV}$.  
%RSG 30Aug05: 0.830 \pm 0.218 \pm 0.081 \pm 0.187
%RSG 01Sep05: removed this sentence as well as the sub-section on it
%We have
%also presented the distribution of $m_{\pi\pi}$, although cannot
%draw meaningful conclusions from it about the production process.

% CURRENT acknowledgments go here...
We gratefully acknowledge the effort of the CESR staff
in providing us with excellent luminosity and running conditions.
This work was supported by the National Science Foundation
and the U.S. Department of Energy.

% The bibliography

\end{document}